\documentclass[11pt,twoside]{article}

\usepackage{asp2014}
\usepackage{xcolor}                  % Good fonts

\aspSuppressVolSlug
\resetcounters

\begin{document}

\title{CASA 6: Modular Integration in Python}

\author{Ryan~Raba$^{1}$, Darrell Schiebel$^{1}$, Bjorn Emonts$^{1}$, Robert Garwood$^{1}$, Federico Montesino Pouzols$^{2}$, Sandra Castro$^{2}$, C. Enrique Garc\'{i}a-Dab\'{o}$^{2}$, David Mehringer$^{1}$, Ville Suoranta$^{1}$}
  \affil{$^1$National Radio Astronomy Observatory, 520 Edgemont Road, Charlottesville, VA 22903}
  \affil{$^{2}$European Southern Observatory, Karl Schwarzschild Strasse 2, D-85748 Garching, Germany}

\begin{abstract}
CASA, the Common Astronomy Software Applications, is the primary data processing software for the Atacama Large Millimeter/submillimeter Array (ALMA) and the Karl G. Jansky Very Large Array (VLA), and is often used also for other radio telescopes. CASA has always been distributed as a single, integrated application, including a Python interpreter and all the libraries, packages and modules. As part of the ongoing development of CASA 6, and the switch from Python 2 to 3, CASA will provide greater flexibility for users to integrate CASA into existing Python workflows by using a modular architecture and standard pip wheel installation. These proceedings of the 2019 Astronomical Data Analysis Software $\&$ Systems (ADASS) conference will give an overview of the CASA 6 project.
\end{abstract}

% These lines show examples of subject index entries. At this stage these have to commented
% out, and need to be on separate lines. Eventually, they will be automatically uncommented
% and used to generate entries in the Subject Index at the end of the Proceedings volume.
% Don't leave these in! - replace them with ones relevant to your paper.
%\ssindex{FOOBAR!conference!ADASS 2019}
%\ssindex{FOOBAR!organisations!ASP}

% These lines show examples of ASCL index entries. At this stage these have to commented
% out, and need to be on separate lines. Eventually, they will be automatically uncommented
% and used to generate entries in the ASCL Index at the end of the Proceedings volume.
% The ascl.py command will scan your paper on possible code names.
% Don't leave these in! - replace them with ones relevant to your paper.
%\ooindex{FOOBAR, ascl:1101.010}

\section{Introduction}

The Common Astronomy Software Applications package CASA \citep[][CASA team et al. in prep.]{mcm07} is the data processing software used by ALMA and the VLA, and supports the various pipelines for ALMA, VLA and VLA Sky Survey (VLASS). CASA has a versatility that also benefits the reduction and imaging of data from other interferometric and single-dish radio telescopes.

The CASA infrastructure historically consisted of a set of C++ tools
bundled together under an iPython interface as data reduction tasks that
are scriptable or can be called via task interface. CASA has always been
distributed as a single integrated application, but many users find it
difficult to use CASA tools and tasks along with the other python
packages. These proceedings describe how users will be able to integrate CASA version 6 into their existing Python environment, with tools and tasks as standard Python modules.

\paragraph{CASA $\&$ Python}

The first release of CASA was in the fall of 2009 using Python 2.5.2. In the intervening years, CASA has upgraded Python a number of times, through 2.6 versions and 2.7 versions. In 2020, the Python community will no longer support Python 2 and CASA will move to Python 3, which was not designed to be backward compatible.

\section{Common Astronomical Software Applications - Version 6}

CASA 6 makes the transition from Python 2.7 to Python 3.6, by migrating the task and tool wrappers and overall execution environment. This involves both developer maintained code and the automated C++ to Python bindings. The transition to Python 3 provides an important motivation for making CASA more Pythonic.

From CASA 6 onward, CASA should be thought of as the system described
in Fig\,\ref{fig:casa6}. CASA tasks and tools will be available as independent Python 3 modules built over top of the C++ code base. These modules can be installed via pip wheels. Graphical User Interface (GUI) applications, such as PlotMS, the CASA Viewer, and the Cube Analysis and Rendering Tool for Astronomy (CARTA) run as separate executable processes that communicate with CASA Python via gRPC. Additional Python modules are provided to automatically start and control these external processes.

\articlefigure[width=0.78\textwidth]{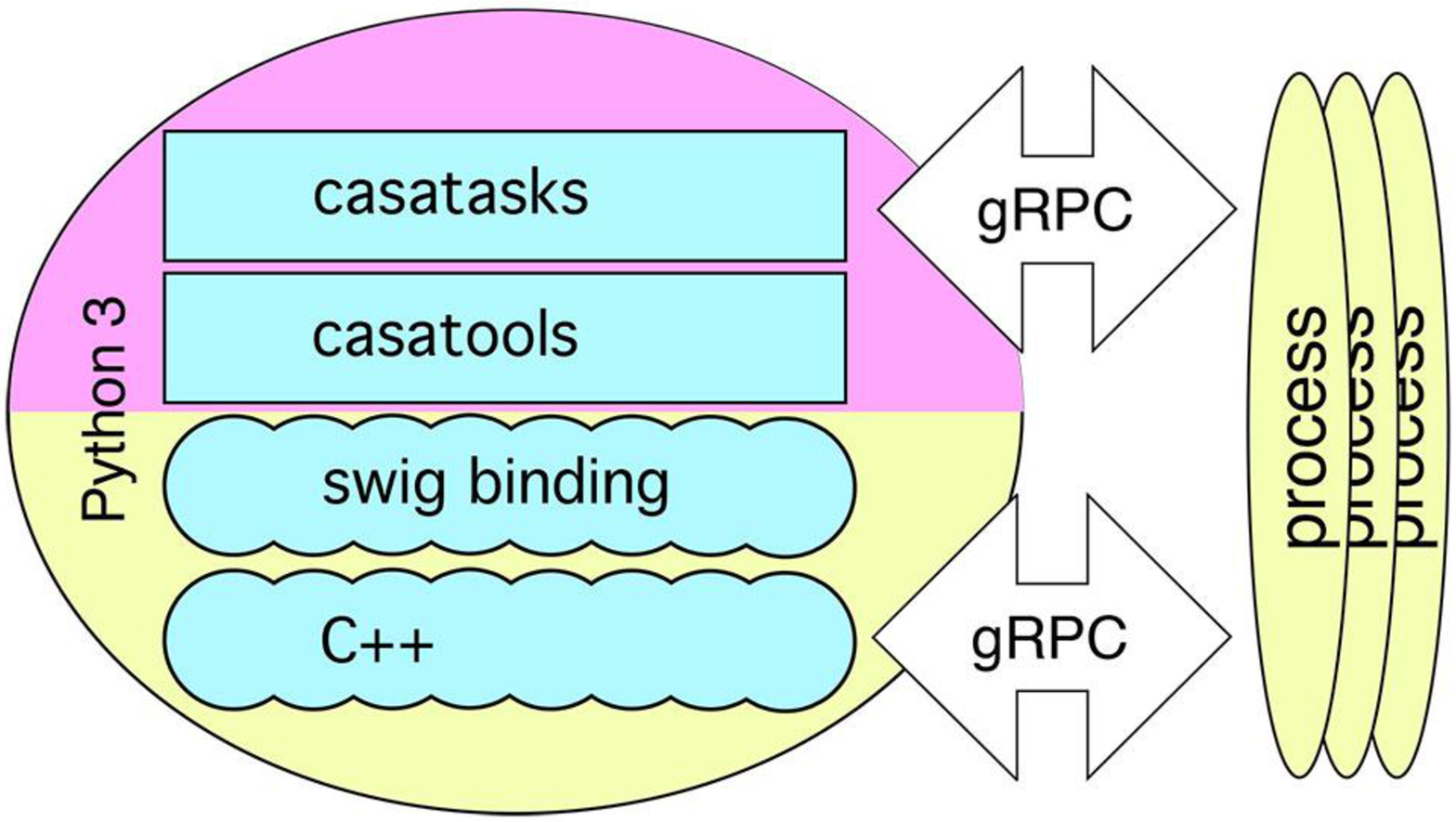}{fig:casa6}{{\bf Schematic overview of CASA 6.} Tasks and tools will be available as independent Python 3 modules that can be installed through pip wheels, while GUIs run as separate executable processes that communicate with CASA Python via gRPC. For example, PlotMS, the CASA Viewer, and CARTA would correspond to the ``process'' ellipses in the diagram. }

\section{Pip wheel installation}

Pip wheels for casatools and casatasks are available as beta-versions
from the public PyPI server casa-pip.nrao.edu. This allows simple
installation and import in to standard Python 3.6 environments. With the pip installation, CASA may be used in a
standard Pythonic manner such as:\\
\ \\
\texttt{from casatasks import listobs}\\
\ \\
\texttt{rc = listobs('mydata.ms')}\\
\ \\
For comprehensive instructions on installation and usage of the initial CASA 6.0 version, see CASA Docs:\\
\textcolor{blue}{https://casa.nrao.edu/casadocs/casa-5.6.0/introduction/casa6-installation-and-usage}

\section{Jupyter Notebook}
\label{sec:jupyter}
Jupyter notebooks are ideally suited for code tutorials, exploration,
and collaborative development. Together with Google Colaboratory, which
hosts Jupyter notebooks on free virtual hardware in the cloud, the door
is opened to powerful new ways of developing and sharing software. CASA
6 casatools and casatasks modules are compatible with the Google Colab
environment. 

The CASA team is working towards making additional modules
compatible in the future as well as introducing new Jupyter-based
CASAguide tutorials. An example of a Jupyter notebook that explains installation and usage of CASA 6 is available here: \textcolor{blue}{https://go.nrao.edu/casa6}

\section{Parallel processing}

CASA adopts so-called embarrassingly parallelization, where a problem can be separated into parallel tasks that need little dependency or communication between them \citep{cas17}. One can parallelize the work by having non-trivially parallelized algorithms that use several processors within a single CASA instance, or by partitioning the MeasurementSet data into a Multi-MS and run a CASA instance on each part. 
%These parallelized data can be processed using the Message Passing Interface (MPI) as a standard which addresses primarily the message-passing parallel programming model in a practical, portable, efficient and flexible way.

Casampi is a Python package that provides the Message Passing Interface (MPI) parallelization infrastructure of CASA 6. It can be used together with casatools and casatasks when using them as separate Python modules in custom user Python setups. Casampi can be used in two scenarios:

\leftskip=1cm
%\vspace{-3mm}\\
\noindent 1). As a plugin for casatasks.\\
\vspace{-4mm}\\
2). As a module that can be imported in order to use the casampi parallelization functionality directly. This is mostly useful for advanced users, such as the ALMA and VLA data reduction pipelines.

\leftskip=0cm
\vspace{1mm}
\noindent In scenario 1, casampi adds parallelization at the CASA tasks level. If casampi is installed together with casatasks, casatasks will see that casampi is available and enable the parallel processing capabilities of CASA. This includes:

\leftskip=1cm
\noindent a) Parallel imaging, following the parallelization approach implemented in the {\sc tclean} task. This is used in ALMA operations since Cycle 6. \\
b) Multi-MS parallelization of various tasks, including flagdata, setjy, etc.

\leftskip=0cm
\vspace{1mm}
\noindent Reports on performance in different setups are available from the official CASA Docs documentation and CASA Memo Series.\footnote{See https://casa.nrao.edu/casadocs/ for CASA Docs and Memo Series.}

\section{Monolithic version}

For users that are not interested in these new modular features, CASA
will continue to offer an all-inclusive distribution under Python 3 as a
single tar file download, starting with CASA 6.1. This monolithic package is built from the same
modular pip wheels, but adds a CASA-shell component to replicate the
appearance of previous CASA versions. This means that the monolithic version will keep practical changes for users to a minimum compared to earlier CASA versions. For the time being, the ALMA and VLA pipeline will also be built on this monolithic version of CASA 6.

\section{Concluding remarks}

Pip wheels for casatools and casatasks will be available as part of CASA 6.0, which was released on 17 Dec 2019. There may be compatibility issues as we refine the included shared libraries to maximize OS support. 

In addition to making CASA more Pythonic in the way that it is accessible within Python, there will also be some more cleanup of problems that have accumulated over the last decade of development with Python. 

All of this means that a period of change is ahead for both CASA and CASA users. To ease the transition, CASA will support two versions for one year. One version will be a single monolithic application based on Python 2.7. The other version, CASA 6.0, will be based on Python 3.\\
\vspace{-2mm}\\
Users are welcome to send any feedback on CASA 6 to {\bf casa-feedback{@}nrao.edu}

\vspace{8mm}

\acknowledgements The CASA team thanks the organizers of the ADASS 2019. CASA is developed by an international consortium of scientists based at the National Radio Astronomical Observatory (NRAO), the European Southern Observatory (ESO), the National Astronomical Observatory of Japan (NAOJ), the Academia Sinica Institute of Astronomy and Astrophysics (ASIAA), the CSIRO division for Astronomy and Space Science (CASS), and the Netherlands Institute for Radio Astronomy (ASTRON), under the guidance of NRAO. The National Radio Astronomy Observatory is a facility of the National Science Foundation operated under cooperative agreement by Associated Universities, Inc. ALMA is a partnership of ESO (representing its member states), NSF (USA) and NINS (Japan), together with NRC (Canada), NSC and ASIAA (Taiwan), and KASI (Republic of Korea), in cooperation with the Republic of Chile. The Joint ALMA Observatory is operated by ESO, AUI/NRAO and NAOJ.  

%\begin{thebibliography}

%\bibitem[Castro et al.(2017)]{cas17} Castro, S., Gonzalez, J., Taylor, J., et al.\ 2017, ADASS XXV, 595

%\bibitem[McMullin et al.(2007)]{mcc07} McMullin, J.~P., Waters, B., Schiebel, D., Young, W., Golap, K.\ 2007, ADASS XVI, 376, 127 

%\end{thebibliography}

\bibliography{P5-6_astroph.bib}  % For BibTex

% if we have space left, we might add a conference photograph here. Leave commented for now.
% \bookpartphoto[width=1.0\textwidth]{foobar.eps}{FooBar Photo (Photo: Any Photographer)}

\articlefigure[width=\textwidth]{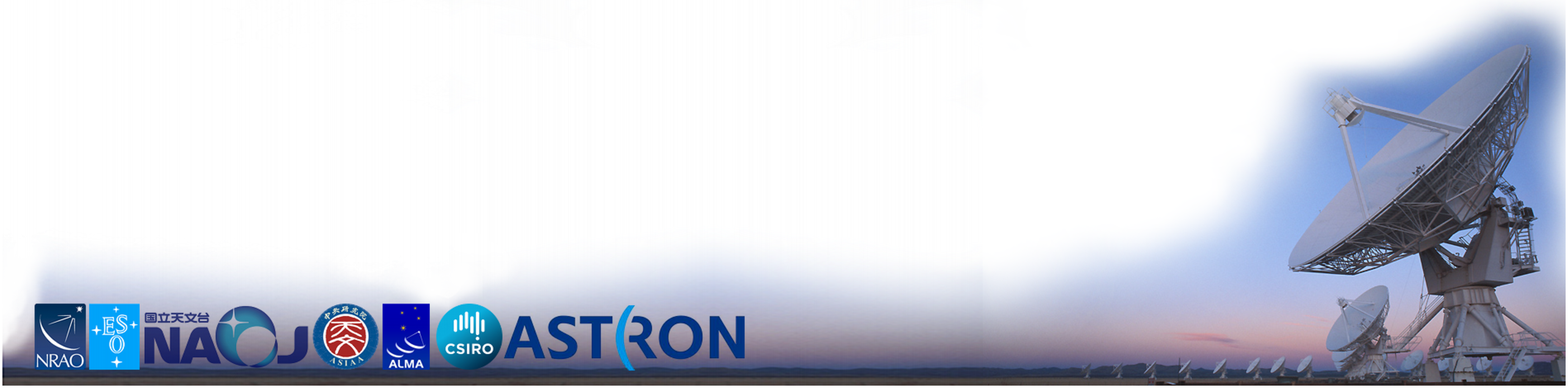}{logo}{}

\end{document}